\documentstyle[eqsecnum,aps,psfig]{revtex}

\def\etal{{\it et\thinspace al.}\ }
\def\dfe{{$\frac{df}{d\epsilon}$}\ }
\def\eion{{(e~+~ion)}\ }
\def\fbar{{$\bar{f}_r$}\ }
\begin{document}
\draft
\preprint{HEP/123-qed}
\title{K-shell dielectronic resonances in photoabsorption: differential
oscillator strengths for Li-like C IV, O VI, and Fe XXIV}
\author{Sultana N. Nahar and Anil K. Pradhan}
\address{
Department of Astronomy, The Ohio State University, Columbus, Ohio
43210\\
}
\author{Hong Lin Zhang}
\address{
Applied Theoretical and Computational Physics Division, Los Alamos
national Laboratory, Los Alamos, NM 87544
}
\date{\today}
\maketitle
\begin{abstract}
 Recently X-ray photoabsorption in KLL resonances of O~VI was 
predicted [Pradhan, Astrophys.J. Lett.
{\bf 545}, L165 (2000)], and detected by the {\it
Chandra X-ray Observatory} [Lee \etal, Astrophys. J. {\it Lett.}, in
press]. The required resonance oscillator
strengths, $\bar{f}_r$, are evaluated 
in terms of the differential oscillator strength
d$f$/d$\epsilon$ that relates bound and continuum absorption. We present
the \fbar values from radiatively damped and undamped
photoionization cross sections for Li-like C,O, and Fe 
calculated using relativistic close 
coupling Breit-Pauli R-matrix method. The KLL resonances of
interest here are: $1s2p (^3P^o) 2s \ \  [^4P^o_{1/2,3/2},
^2P^o_{1/2,3/2}]$ and $1s2p (^1P^o) 2s \ \ [^2P^o_{1/2,3/2}]$.
The KLL photoabsorption resonances in Fe~XXIV are fully resolved up to
natural autoionization profiles for the first time.
It is demonstrated that the undamped $\bar{f}_r$ independently
yield the resonance radiative decay rates, and thereby provide a precise
check on the resolution of photoionization calculations in general. 
The predicted photoabsorption features should be detectable by the X-ray space
observatories and enable column densities in highly ionized astrophysical
plasmas to be determined from the calculated $\bar{f}_r$. The
dielectronic satellites may appear as redward broadening 
of resonance lines in emission and absorption.

\end{abstract}


Resonance spectroscopy may be employed to determine element abundances and
ionization structure in astrophysical sources \cite{p00,letal}.
 K-shell line spectra are powerful diagnostic
tools in X-ray spectroscopy for density, temperature, ionization state, 
and abundances laboratory and astrophysical plasmas
(e.g. \cite{setal,metal}). K-shell
emission spectra from several atomic species have been measured in
Electron-Beam-Ion-Trap experiments \cite{setal,bei92}. Recently it was
suggested that, analogous to the dielectronic satellite
spectra of He-like ions in emission, K-shell X-ray photoabsorption 
should be detectable in KLL resonances of Li-like O~VI \cite{p00}.
Soon thereafter, the {\it Chandra
X-ray Observatory} (CXO) made the first detection of the strongest of the
predicted O~VI KLL features at 22.05 $\AA$ from
the so called `warm absorber' region of ionized gas around the 
central source in active galactic nuclei, thought to be a massive 
black hole \cite{letal}. 
The observation of X-ray absorption from Li-like O~VI in the
same wavelength region as the emission lines of He-like O~VII 
enabled the simultaneous
determination of column densities and, since the two ions do not
usually co-exist in a plasma, indicated a composite plasma with
widely disparate temperature regimes \cite{p00,letal}.

 Present theoretical calculations for the resonance oscillator strengths
employ highly resolved relativistic photoionization cross sections
with fine structure \cite{zp97,p00}. 
The quantity
of interest is the differential oscillator strength \dfe that quantitatively
relates photoabsorption per unit energy in the bound-bound and the
continuum region as \cite{s83,fr86,ps77}
\begin{equation}
\frac{df}{d\epsilon} = \left[ \begin{array}{l}
                         \frac{\nu^3}{2z^2} f_{\rm line} \ \ \ \ ,
\hskip 2cm \epsilon < I \\
                         \frac{1}{4\pi^2 \alpha a_0^2} \sigma_{\rm PI}
 \ , \hskip 2cm \epsilon > I
                         \end{array}
                         \right.
\end{equation}

\noindent where $f_{\rm line}$ is the line absorption oscillator strength,
$\sigma_{\rm PI}$ the photoionization cross section, $I$ the ionization
potential, $z$ the ion charge,
$\nu$ the effective quantum number
at $\epsilon = -\frac{z^2}{\nu^2}$ in Rydbergs,
and  $\alpha$ and $a_0$ are the fine structure constant
and the Bohr radius respectively. In the bound-free
region, with autoionizing resonances,  the integrated \dfe yields the
effective photoabsorption in terms of $\sigma_{\rm PI}$, i.e.

\begin{equation}
 \bar{f}_{r} (J_i \longrightarrow J_f)  =  \int_{\Delta E_{r}}
\left(
\frac{df (J_i \longrightarrow J_f)}{d\epsilon} \right) d\epsilon  
= \left( \frac{1}{4\pi^2 \alpha
a_0^2} \right) \int \sigma_{\rm PI} (\epsilon; J_i \rightarrow J_f) d\epsilon ,
\end{equation}

\noindent where $J_i,J_f$ are total angular momenta of the initial 
bound level and the
final bound or continuum wavefunction, governed by the
usual dipole selection rules $\Delta J = 0, \pm 1; \pi \rightarrow -
\pi$. Eq. (0.2) may be evaluated from the detailed $\sigma_{\rm PI}$ for the
$J\pi$ symmetries concerned. However,
the resonance profile needs to be sufficiently well delineated and
elaborate methods need to be
employed to obtain accurate positions and profiles (the background and
the peaks) of resonances.
Relativistic effects are essential in order to differentiate the fine
structure components. We employ the coupled channel formulation based
on the relativistic Breit-Pauli R-matrix (BPRM) method
as employed on the extensive work on the Iron Project 
\cite{ip,betal,ipwww}.

 BPRM photoionization and recombination calculations
were reported in earlier works for Li-, He-, H-like carbon and iron: 
C~IV, C~V, C~VI \cite{zetal,n00a}
and Fe~XXIV, Fe~XXV, Fe~XXVI \cite{pz97,n00b} for applications to X-ray
photoionization and non-LTE (NLTE) modeling.
We consider the photoionization of the ground level of the Li-like ions
C~IV, O~VI, and Fe~XXIV, $1s^22s \ (^2S_{1/2})$ into
all $n=1,2,3$ fine structure levels of the He-like ion with $1s^2 (^1S_0), 1s2s
(^3S_1,^1S_0)$, $1s2p (^3P^o_{0,1,2},^1P^o_{1})$, $1s3s (^3S_1,^1S_0),
1s3p (^3P^o_{0,1,2},^1P^o_1)$ and
$1s3d (^3D_{1,2,3}, ^1D_2)$ in the close coupling
expansion. Thus K-shell photoionization of the $(1s^22s)$, i.e.
inner-shell excitation-autoionization
via the $1s \rightarrow 2p$ transition resulting in $1s2s2p$ (KLL)
resonances, is considered with
the initial bound state ($^2S_{1/2}$) with symmetry $J = 0.5$ (even parity), 
and final continua
with $J = 0.5$ and 1.5 (odd parity). The six KLL resonances of
interest here are: $1s2p (^3P^o) 2s \ \  [^4P^o_{1/2,3/2},
^2P^o_{1/2,3/2}]$
and $1s2p (^1P^o) 2s \ \ [^2P^o_{1/2,3/2}]$ (the resonances are 
labeled in Table 1 according to the alphabetical
notation by Gabriel \cite{g72}). The autoionization and
radiative decay rates and cross sections, with and without radiative
decay of resonances back to the ground level, are calculated by analyzing 
the poles in the complex dipole matrix elements using the method described in 
\cite{pz97,saki}. The cross sections are resolved on a very fine energy
mesh of up to $10^{-6}$ eV.

 Whereas calculations have been carried out for C~IV, O~VI, and Fe~XXIV,
for brevity we present only the detailed photoionization cross section 
of the Fe~XXIV ground state in Fig.~1 from the
L-shell ($2s$) ionization threshold at Fe~XXV $(1s^2 \ ^1S_0)$, up to the
K-shell ionization thresholds at $1s2s$ and $1s2p$ (marked by arrows in the
figure). The top panel (a) presents the $\sigma_{\rm PI}$ over the extended 
range
with the KL$n\ n \geq 2$ complexes of resonances KLL, KLM, KLN, etc. 
(at computed wavelengths) converging on to the K-shell ionization edges. 
The bottom panel (b) shows the region of KLL resonance complex 
on an expanded scale, and the strongest resonance (labeled `q' , Table
1), fully delineated in (c), showing the orders of magnitude effect
of radiation damping. The calculations are further
subdivided into the constituent total $J\pi$ symmetries for the \eion
continua, and we thereby fully
resolve the lowest KLL resonance complex into the six distinct resonance
features allowed by the orbital and spin angular momentum couplings, 
as shown in the bottom panel of Fig. 1. Two sets of cross sections are computed
-- with and without radiation damping of resonances (solid and dashed
lines respectively). It is noted that the Y-scale (Megabarns) in the upper
panel is linear, whereas in the bottom panel it is log$_{10}$, i.e. the
resonance peaks rise orders of magnitude above the nonresonant background.
Further, the radiative damping effect 
attenuates the peak resonance strengths by similarly large amounts, but
it is different for each resonance. 

  These resonances  have heretofore been observed in the laboratory as
dielectronic satellites (DES) formed via electron-ion excitation of the
core ion followed by dielectronic recombination (DR), i.e. radiation
damping of the core of the doubly excited autoionization state (e.g
Li-like C~IV [4]). We may distinguish the DR resonances DES,
from the `photoabsorption resonances' (PAR) as:

\begin{equation}
\begin{array}{llll}
e + 1s^2 \longrightarrow & 1s2s2p \longrightarrow & 1s^2 \ 2s + h\nu &
\mbox{(DES)}\\
h\nu + 1s^2 \ 2s & \longrightarrow 1s2s2p & &  \mbox{(PAR).}
\end{array}
\end{equation}

 The PAR's therefore manifest themselves in photoionization cross
sections and photoabsorption spectra, whereas the DES appear in emission
spectra. For lighter elements such as
C and O, the PAR's are not likely to be also observed as DES. Thus, in
general, resonances may be divided as PAR or DES according to their
actual formation and observation depending on plasma conditions in the
source. For example,
in ions of heavier elements such as Fe~XXIV, we expect the same set of
resonances to act as PAR's {\it or} the DES, and provide
useful plasma diagnostics
 -- absorption from cold or cooling environment (with Li-like
Fe~XXIV for example), or emission at higher temperatures (with DR
from He-like Fe~XXV). Since both the DES and the PAR are satellite
features on the long wavelength side of the resonance lines, they 
should appear as `redward broadening' of the primary
resonance transitions in emission or absorption spectra.
This fact may be of considerable astrophysical significance
in the analysis of X-ray spectra where such asymmetric broadening of
resonance lines is observed, and is sometimes ascribed to relativistic
gravitational broadening due to proximity of the plasma source to
a massive black hole (see the discussion in Ref. \cite{letal}).

Table 1 gives the computed resonance parameters: the associated
autoionization and radiative rates ($\Gamma_a,\Gamma_r$), 
as well as the radiatively undamped (NRD) and the damped (RD)
\fbar's for the KLL resonances in C~IV, O~VI, and Fe~XXIV.
The \fbar (NRD) are the conventional $f$-values, but computed for
resonances; the \fbar (RD) are further multipled by the ratio of the
radiative rate back to the ground level and the sum of all radiative
rates.
The \fbar (C~IV,NRD) agree well with the length $f$-values computed in 
\cite{metal} to 3.9\% for the $1s2p(^1P^o)2s$, and to 0.9\% for the 
$1s2p(^3P^o)2s$ resonances. 

The `q' resonance in Fe~XXIV (Fig. 1c) is the narrowest one with an extremely
small $\Gamma_a$, the lowest among the KLL PAR's. 
But it has the largest $\Gamma_r$, and 
hence the largest resonance oscillator strength \fbar, resulting in the most 
damped profile.
In previous works the resonance $\Gamma_a,\Gamma_r$ are generally determined 
either in the isolated resonance approximation (e.g. \cite{vs78,bdetal}),
or through fitting of complex dipole matrix elements for the
dielectronic satellite resonances (e.g. \cite{saki,zetal,pz97}). The
resonances themselves are not fully
resolved up to their natural autoionization widths.
In the present work, the `q' resonance required an order of magnitude
finer resolution than that needed to derive $\Gamma_a,\Gamma_r$ from
fitting alone (given in Table 1). The \fbar's obtained from integration over
completedly resolved \dfe accounting for all the oscillator strength
(as for Fe~XXIV), should be somewhat more accurate than
from the fits, or isolated resonance approximation when
interference between the background and the resonant parts of the
wavefunctions is significant. The \fbar in Table 1 agree very well 
with those derived from fits to resonances \cite{pz97}.
Another independent check is that the
sum over the \fbar's for the KLL resonances of Fe~XXIV, 0.782, 
agrees well with the total absorption 
oscillator strength in the E1 dipole allowed and intercombination
transitions in the core ion Fe~XXV $f \ (1s^2 \ ^1S_0
- 1s2p \ ^{(3,1)}P^o_1)$, 0.772. 

The present calculations
therefore enable not only the calculation of PAR oscillator
strengths, but also provide the most stringent check on the accuracy of
photoionization calculations.
 For  C~IV and O~VI all six KLL resonances are not fully delineated due
to overlapping and weak components, but the
stronger components account for nearly all the expected oscillator
strength; the weak components have very small $\Gamma_a$ {\it and} small \fbar.
We should expect the KLL resonances in C~IV and O~VI to manifest 
themselves only as PAR's, and not DES, since the radiative rates are
small compared to autoionization rates. 

The computed \dfe for the three ions are shown in Fig. 2,
eextending from the bound-bound region
with the ($2s - np$) transitions to the bound-free ($2s-\epsilon p$) in
the X-ray region. It may be noted that the bound-bound region is in the
UV for C~IV and O~VI, but in the X-ray for Fe~XXIV with the exception
of the $n$ = 2 transition in the extreme UV.
The bound-bound oscillator strengths are computed up to $n$ = 10 from the same
eigenfunction expansion as in the photoionization and recombination work
in Refs.~\cite{n00a,n00b}; all atomic parameters are thereby obtained in a
theoretically self consistent manner.
The Fe~XXIV and Fe~XXV oscillator strengths were reported in Ref.~\cite{np99} 
(those for carbon and oxygen ions are as yet
unpublished but may be obtained from the first author).
The \dfe are continuous across the ionization threshold, but reflect
magnitudes and the structures of discrete and resonant transitions
\footnote{Owing to a typographical error in the plotting routine, 
$\frac{df}{d\epsilon} (O~VI, n=2)$ = 0.02115 is incorrectly shown 
in Fig.2 of Ref.~\cite{p00} as 0.2115.}.  In the
bound-bound region the  \dfe $(2s - np)$ is much smaller for $n$ = 2 than
for  $n$ = 3, owing to the respective energy differences. The
\dfe therefore rises from $n$ = 2 to 3, and then
decreases monotonically towards higher $n$ and the first ionization 
threshold as $n \rightarrow \infty$ ($\epsilon \rightarrow 0$). 
Fig. 2 presents a complete picture of photoabsorption from the Li-like
C~IV, O~VI and Fe~XXIV, from L-shell absorption to the 
the KLL X-ray features, that should provide plasma diagnostics
in two quite different wavelength ranges {\it from the same ionic species}.

The wavelengths shown in Fig.~2, together with the \fbar in Table 1, mark the
positions and strengths of discrete photoabsorption in the
X-ray by Li-like C,O, and Fe.  The predicted features and the calculated
resonance oscillator strengths \fbar should be useful
in the interpretation of X-ray spectra from the CXO and the {\it X-ray
Multi-Mirror Mission}, and determination of column densities using
the standard curve-of-growth method \cite{sp78}. Further, the
significant differences between \fbar (NRD) and \fbar (RD), without and
with radiative damping, may indicate optical depth effects in the
plasma; \fbar (NRD) is the value to be used in optically thin cases
where re-emission following photoabsorption in the resonance along the line 
of sight is unlikely. More generally, this work demonstrates that
resonance photoabsorption spectroscopy can provide information
complementary to line emission spectroscopy, and should prove to 
be a powerful tool for plasma diagnostics, such as for ionization fractions and
abundances of elements in astrophysical and laboratory plasmas.

 This work was partially supported by the National Science Foundation
and the NASA Astrophysical Theory Program. The computational work was
carried out at the Ohio Supercomputer Center.

\def\amp{{Adv. At. Molec. Phys.}\ }
\def\apj{{ Astrophys. J.}\ }
\def\apjs{{Astrophys. J. Suppl.}\ }
\def\apjl{{Astrophys. J. (Lett.)}\ }
\def\aj{{Astron. J.}\ }
\def\aa{{Astron. Astrophys.}\ }
\def\aasup{{Astron. Astrophys. Suppl.}\ }
\def\adndt{{At. Data Nucl. Data Tables}\ }
\def\cpc{{Comput. Phys. Commun.}\ }
\def\jqsrt{{J. Quant. Spectrosc. Radiat. Transf.}\ }
\def\jpb{{J. Phys. B}\ }
\def\pasp{{Pub. Astron. Soc. Pacific}\ }
\def\mn{{Mon. Not. R. Astron. Soc.}\ }
\def\pra{{Phys. Rev. A}\ }
\def\prl{{Phys. Rev. Lett.}\ }
\def\zpds{{Z. Phys. D Suppl.}\ }
\def\adndt{At. Data Nucl. Data Tables}

\newpage
\widetext

\begin{table}
\caption{X-ray wavelengths and KLL photoabsorption resonance 
oscillator strengths 
for C~IV, O~VI, and Fe XXIV. Notation $a\pm b$ means $a \times 10^{\pm b}$.
The NRD and RD values refer to without and with radiation damping.
$\Gamma_a$ and $\Gamma_r$ are the autoionization and radiative decay rates
respectively.}
\begin{tabular}{llllll}
\multicolumn{1}{c}{Resonance} & \multicolumn{1}{c}{$\lambda_{cal}$} & 
\multicolumn{1}{c}{E$_r$} & \multicolumn{1}{c}{$\bar{f}_r$}
 & \multicolumn{1}{c}{$\Gamma_a$} & \multicolumn{1}{c}{$\Gamma_r$} \\
& $(\AA)$ & \multicolumn{1}{c}{(KeV)} & \multicolumn{1}{c}{NRD (RD)} & 
\multicolumn{1}{c}{(Ry, s$^{-1}$)} & \multicolumn{1}{c}{(Ry, s$^{-1}$)} \\
\hline
\multicolumn{6}{c}{C IV} \\
\hline
$1s2p(^1P^o)2s \ ^2P^o_{1/2}$ (r) & 41.42 & 0.2993 & 0.3402 (0.3148) &
3.24-4, 6.70+12& 2.97-5, 6.13+11\\
$1s2p(^1P^o)2s \ ^2P^o_{3/2}$ (u) & 41.42 & 0.2993 & 0.1694 (0.1571) &
3.18-4, 6.58+12& 2.93-5, 6.05+11\\
$1s2p(^3P^o)2s \ ^2P^o_{1/2}$ (t) & 40.94& 0.3028 & 0.0342 (0.0342) &
2.75-3, 5.69+13& 3.47-6, 7.16+10\\
$1s2p(^3P^o)2s \ ^2P^o_{3/2}$ (s) & 40.94& 0.3028 & 0.0177 (0.0177) &
2.75-3, 5.69+13& 3.34-6, 6.91+10\\
\hline
\multicolumn{6}{c}{O VI} \\
\hline
$1s2p(^1P^o)2s \ ^2P^o_{1/2}$ (r) & 22.05 & 0.5623 & 0.1924 (0.141) & 
3.11-4, 6.42+12& 1.26-4, 2.60+12\\
$1s2p(^3P^o)2s \ ^4P^o_{3/2}$ (u) & 22.05 & 0.5623 & 0.3837 (0.267) &
2.76-4, 5.70+12& 1.28-4, 2.65+12\\
$1s2p(^3P^o)2s \ ^2P^o_{1/2}$ (t) & 21.87 & 0.5670 & 0.0217 (0.0216) &
3.41-3, 7.11+13 & 1.46-5, 3.01+11\\
$1s2p(^3P^o)2s \ ^2P^o_{3/2}$ (s) & 21.87 & 0.5670 & 0.0391 (0.0390) &
3.43-3, 7.09+13&1.32-5, 2.72+11\\
\hline
\multicolumn{6}{c}{Fe XXIV} \\
\hline
$1s2p(^1P^o)2s \ ^2P^o_{1/2}$ (v) & 1.873 & 6.617   & 2.05-3 (1.61-4) &
7.48-6, 1.54+11& 1.77-4, 3.66+12\\
$1s2p(^1P^o)2s \ ^2P^o_{3/2}$ (u) & 1.870 & 6.627   & 1.42-2 (7.63-4) &
3.12-5, 6.45+11& 6.38-4, 1.32+13\\
$1s2p(^1P^o)2s \ ^2P^o_{1/2}$ (r) & 1.864 & 6.649   & 1.46-1 (1.77-2) &
1.89-3, 3.90+13& 1.35-2, 2.80+14\\
$1s2p(^3P^o)2s \ ^4P^o_{3/2}$ (q) & 1.860 & 6.663   & 4.91-1 (2.58-4) &
8.98-6, 1.86+11& 2.29-2, 4.74+14\\
$1s2p(^3P^o)2s \ ^2P^o_{1/2}$ (t) & 1.857 & 6.674   & 1.10-1 (2.66-2) &
3.43-3, 7.10+13& 1.04-2, 2.14+14\\
$1s2p(^3P^o)2s \ ^2P^o_{3/2}$ (s) & 1.855 & 6.681   & 6.56-3 (6.24-3) &
5.12-3, 1.06+14& 2.82-4, 5.83+12\\
\end{tabular}
\end{table}
\newpage
\begin{figure}
\centering
\psfig{figure=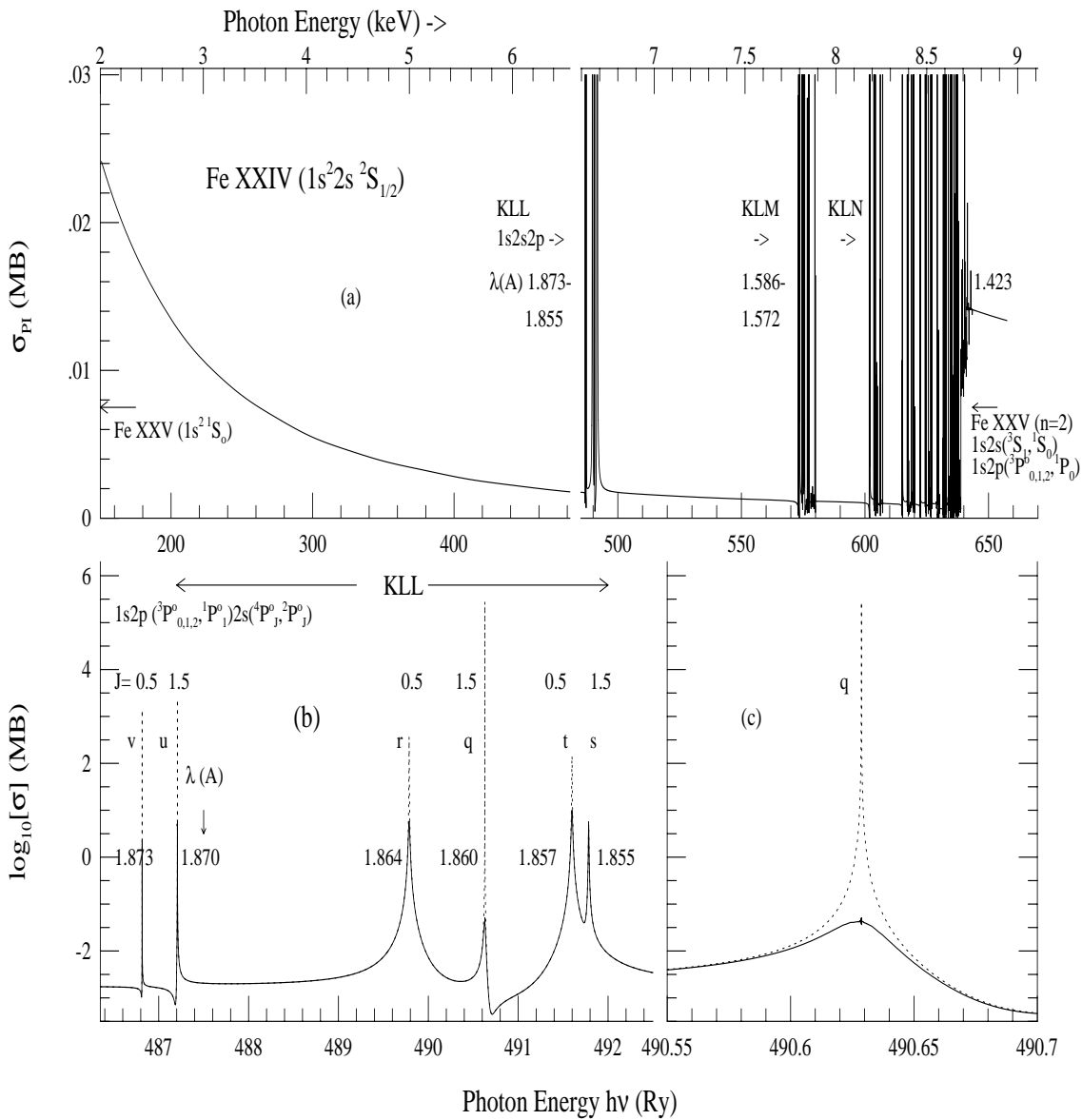,height=17.0cm,width=18.0cm}
\caption{Ground state photoionization cross section of Fe~XXIV with the
K-complex of resonances (a), the resolved
KLL resonances (b), and completely resolved profile of the q-resonance
(c). Note the different energy scales on the X-axis in (a), and in (b) and (c).
The upper panel has a
linear Y-scale (Megabarns), but the lower panel is on a log$_10$ scale
indicating the height of resonances. The large radiation
damping of the `q' resonance (c) is related to its large
\fbar, stronger than all other resonances.}
\end{figure}

\begin{figure}
\centering
\psfig{figure=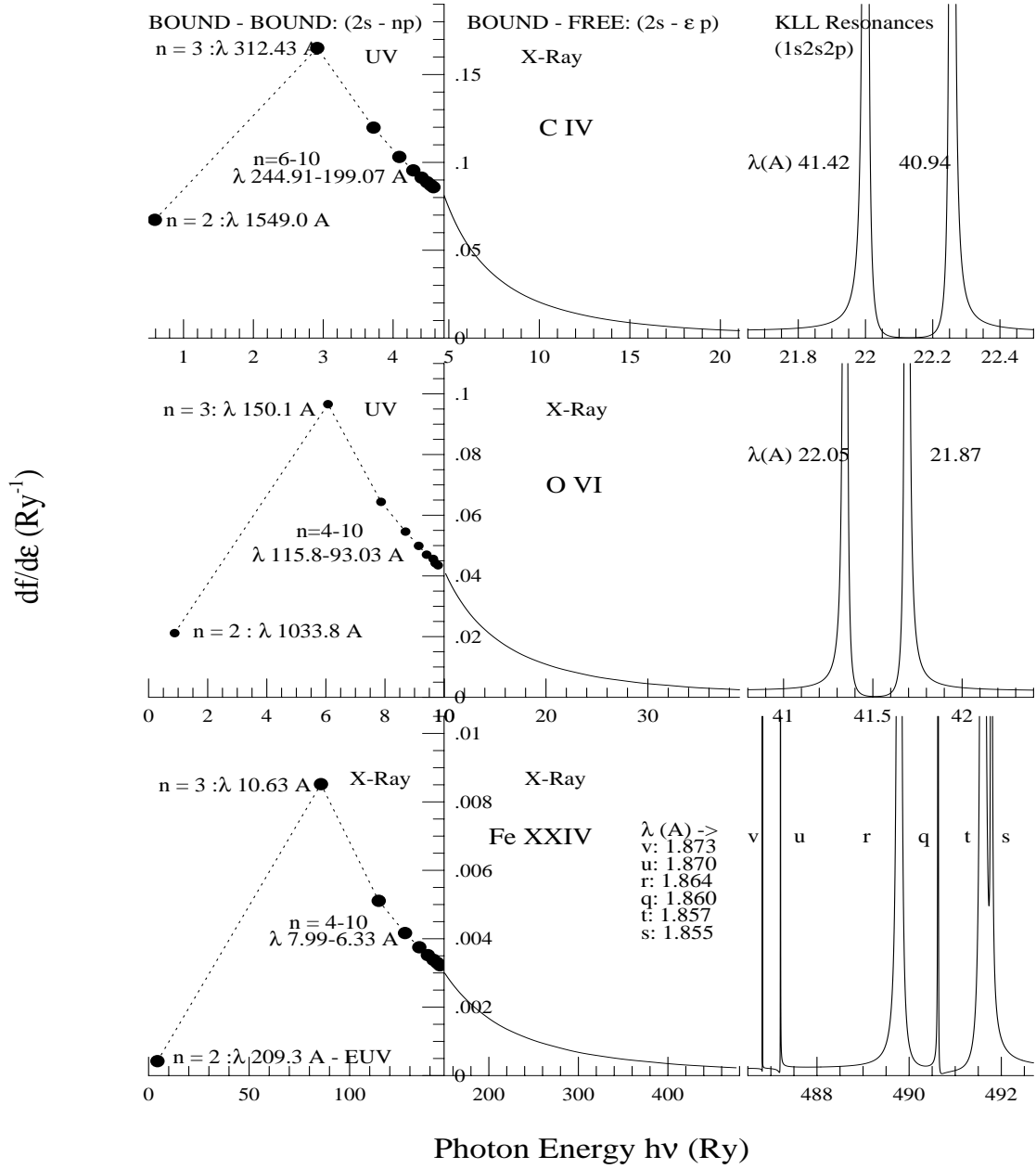,height=20.0cm,width=18.0cm}
\caption{Differential oscillator strengths \dfe for C~IV, O~VI, and Fe~XXIV.
Wavelengths of the main line and resonance features are
marked. The PAR \fbar's in Table 1 give the expected strengths in
X-ray absorption spectra.}
\end{figure}

\widetext


\end{document}